\documentclass[aps, pra, twocolumn, nofootinbib, superscriptaddress]{revtex4-1}
\usepackage{amsfonts}
\usepackage{amsmath}
\usepackage{amssymb}
\usepackage{graphicx}

\usepackage{comment}
\usepackage{color}
\usepackage{cancel}
\usepackage{ulem}

\def\filetype{eps}

\begin{document}
\title{Analytical study of static beyond-Fr\"{o}hlich Bose polarons in one dimension}
\author{Ben Kain}
\affiliation{Department of Physics, College of the Holy Cross, Worcester, Massachusetts 01610 USA}
\author{Hong Y.\ Ling}
\affiliation{Department of Physics and Astronomy, Rowan University, Glassboro, New Jersey 08028 USA}

\begin{abstract}
\noindent
Grusdt et al.\ [New J.\ Phys.\ 19, 103035 (2017)] recently made a renormalization
group study of a one-dimensional Bose polaron in cold atoms. Their
study went beyond the usual Fr\"{o}hlich description, which includes only
single-phonon processes, by including two-phonon processes in which two
phonons are simultaneously absorbed or emitted during impurity scattering
[Shchadilova et.\ al., Phys.\ Rev.\ Lett.\ 117, 113002 (2016)]. We study this same beyond-Fr\"{o}hlich model, but in the static impurity limit where the ground state is described
by a multimode squeezed state instead of the multimode coherent state in the
static Fr\"{o}hlich model. We solve the system exactly by applying the
generalized Bogoliubov transformation, an approach that can be
straightforwardly adapted to higher dimensions. Using our exact solution, we
obtain a polaron energy free of infrared divergences and construct analytically the polaron phase diagram. We find that the repulsive polaron is stable on
the positive side of the impurity-boson interaction but is always
thermodynamically unstable on the negative side of the impurity-boson
interaction, featuring a bound state, whose binding energy we obtain
analytically. We find that the attractive polaron is always dynamically unstable,
featuring a pair of imaginary energies which we obtain analytically. 
We expect the multimode squeezed state to help with studies
that go
not only beyond the Fr\"{o}hlich paradigm but also beyond Bogoliubov theory,
just as the multimode coherent state has helped with 
the study of Fr\"{o}hlich polarons.

\end{abstract}
\maketitle

\section{Introduction}

An impurity submerged in a cold-atom Bose--Einstein condensate (BEC)
represents an open quantum system.
The concept of a Bose polaron emerges naturally in this system as an impurity dressed with phonons, where the phonons are low-energy excitations associated with density fluctuations of the BEC. 
That the impurity-phonon coupling
in a BEC-polaron system may be modeled by terms linear in phonon fields makes
the BEC-polaron system the cold-atom analog of the Fr\"{o}hlich model for the
electron-phonon system
\cite{landau46ZhEkspTeorFiz.16.341,landau48ZhEkspTeorFiz.18.419,frohlich54AdvPhys.3.325}. 
The Fr\"{o}hlich paradigm has been quite influential in the study of many exciting phenomena,
including high-temperature superconductivity in solid-state systems (see
\cite{alexandrov07Book} for a review).
Of particular interest is the
Fr\"{o}hlich polaron in the regime 
with strong coupling between the
impurity and phonons. In solid-state systems, the electron-phonon coupling
is fixed by the underlying crystal structure. This limitation makes the
Fr\"{o}hlich polaron in the strong-coupling regime virtually inaccessible to
solid-state experiments, despite extensive efforts in improving our
theoretical understanding of such polarons
\cite{landau46ZhEkspTeorFiz.16.341,landau48ZhEkspTeorFiz.18.419,lee53PhysRev.90.297,feynman55PhysRev.97.660}.
In contrast, in cold-atom systems the impurity-phonon coupling can be
made arbitrary large by tuning the interspecies interaction across a Feshbach
resonance \cite{chin10RevModPhys.82.1225}.  Cold atoms, then, are an
excellent platform for exploring strongly interacting polarons (for recent
reviews, see \cite{devreese13arXiv:1012.4576,grusdt15arXiv:1510.04934}).
As such, there has been a
flurry of theoretical
\cite{astrakharchik04PhysRevA.70.013608,cucchietti06PhysRevLett.96.210401,kalas06PhysRevA.73.043608,bruderer08NewJournalOfPhysics.10.033015,
Huang09ChinesePhysicsLetters.26.080302,tempere09PhysRevB.80.184504,casteels11LaserPhysics.21.1480,casteels12PhysRevA.86.043614,rath13PhysRevA.88.053632,kain14PhysRevA.89.023612,
shashi14PhysRevA.89.053617,li14PhysRevA.90.013618,grusdt15ScientificReports.5.12124,ardila15PhysRevA.92.033612,vlietinck15NewJournalOfPhysics.17.033023,shchadilova16PhysRevA.93.043606,
sogaard15PhysRevLett.115.160401,ardila16PhysRevA.94.063640,levinsen15PhysRevLett.115.125302,kain16PhysRevA.94.013621,
sogaard15PhysRevLett.115.160401,levinsen15PhysRevLett.115.125302,dehkharghani15PhysRevA.92.031601,ardila16PhysRevA.94.063640,kain16PhysRevA.94.013621,
schmidt16PhysRevLett.116.105302,parisi17PhysRevA.95.023619,sun17PhysRevLett.119.013401,volosniev17PhysRevA.96.031601,yoshida18PhysRevX.8.011024, arXiv:1807.09992, arXiv:1807.09948}
and experimental
\cite{catani12PhysRevA.85.023623,hu16PhysRevLett.117.055301,jorgensen16PhysRevLett.117.055302,meinert17Science356.945}
activity devoted to the subject of Bose polarons in cold atoms.  

Much of the theoretical work in recent years, however, has been done within the
framework of the Fr\"{o}hlich paradigm 
\cite{tempere09PhysRevB.80.184504,casteels11LaserPhysics.21.1480,casteels12PhysRevA.86.043614,vlietinck15NewJournalOfPhysics.17.033023,grusdt15ScientificReports.5.12124,shchadilova16PhysRevA.93.043606,kain16PhysRevA.94.013621}, {and is therefore only valid when linearity is maintained in the phonon field operators for the impurity-phonon coupling.
In crystal lattices, this linear relationship comes about because the
electron's potential energy is proportional to the displacement of ions from
their equilibrium positions and effects from anharmonicity are usually negligible
\cite{feynman90StatisticalMechanicsBook}.  In contrast, the linearity in
atomic BECs comes about because of impurity scattering of bosons between the
condensed and the noncondensed modes.  However, the impurity can also
scatter bosons just between noncondensed modes. As pointed out recently by
Shchadilova et al.\ \cite{shchadilova16PhysRevLett.117.113002}, when the
impurity-boson interaction is strong, an accurate description requires
including such scattering, which, because both modes are to be treated quantum
mechanically, introduces terms bilinear in phonon field operators, leading to
a polaron model that goes beyond the Fr\"{o}hlich paradigm
\cite{shchadilova16PhysRevLett.117.113002,grusdt17PhysRevA.96.013607,grusdt17NewJournalOfPhysics.19.103035}.

Of special relevance to our work here is a recent paper by Grusdt et al.\
\cite{grusdt17NewJournalOfPhysics.19.103035} on Bose polarons in one dimension,
which analyzed the experiment by Catani et al.\
\cite{catani12PhysRevA.85.023623} using a beyond-Fr\"{o}hlich model where
bosons are described within Bogoliubov theory. Despite the significant
simplification afforded by Bogoliubov theory, as long as the impurity
remains mobile, there is no known way to exactly solve the system (with
arbitrary impurity-boson coupling and in the thermodynamic limit), be it
modeled by the usual or the beyond-Fr\"{o}hlich Hamiltonian. 
We base our study on the same model in the paper by Grusdt et al.\
\cite{grusdt17NewJournalOfPhysics.19.103035}, except we treat the impurity as
static (i.e.\ localized in space). 
We carry out a field theoretical analysis of this static model, which is exactly solvable, irrespective of the impurity-phonon coupling strength.  We expect such an exact treatment to offer insights that are valuable to ongoing efforts of constructing many-body field theoretic descriptions of (mobile) Bose polarons which go not only beyond the Fr\"{o}hlich paradigm but also beyond Bogoliubov theory.
(The term ``polaron" is usually reserved for a dressed mobile impurity.  Since the static impurity can be viewed as a mobile impurity in the heavy-mass limit, throughout this work we continue to use the same term for a dressed static impurity.)

Our paper is organized as follows. In Secs.\ II and III, we briefly review
the model and mean-field solution. In Sec.\ IV, following a translation to
displace phonon fields by their mean-field values, we apply the generalized
Bogoliubov transformation to diagonalize (and hence solve exactly) the
Hamiltonian associated with quantum fluctuations around the mean-field
solution. It has been well established that observables in one-dimensional (1D) systems are
plagued by infrared (IR) divergences which trace their origin to a dramatic
increase in the density of states relative to its higher-dimensional
counterparts. The polaron energy obtained using our exact solution is
automatically free of the IR divergence that Grusdt et al.\
\cite{grusdt17NewJournalOfPhysics.19.103035} managed to eliminate from the
mean-field polaron energy using renormalization-group flow equations. In Sec.\
V, we investigate in detail the eigenvalues and corresponding eigenvectors of
the Bogoliubov--de Gennes (BDG) equation for quantum fluctuations and construct the
polaron phase diagram analytically. In agreement with Grusdt et al.\
\cite{grusdt17NewJournalOfPhysics.19.103035}, we find that the repulsive polaron on
the attractive side of the impurity-boson interaction is distinguished by a bound state.  Further, we obtain an analytical expression for the
binding energy of this bound state. For the attractive polaron, quantum Monte
Carlo simulations by Parisi and Giorgini \cite{parisi17PhysRevA.95.023619}
suggest that for weak boson-boson repulsion, bosons may undergo an instability
towards collapse around the impurity. For our static case, we show that the
attractive polaron branch is always dynamically unstable within Bogoliubov
theory and further we provide an analytical formula for determining the rate
at which perturbations grow. We conclude our study in Sec.\ VI.


\section{Model and Hamiltonian}

We consider a cold atom mixture with an extreme population imbalance where
minority atoms are so outnumbered by majority atoms (bosons) that they can be
considered impurities submerged in a bath of bosons. We assume that the two
species have sufficiently different polarizabilities that impurities and host
bosons can be independently manipulated by optical lattices. We specialize
to the situation where one optical lattice traps and localizes impurities
while another optical lattice confines host bosons to a quasi-1D geometry (a
tube).  In a nutshell, we base our theory on the experimental set-up
described by Knap et al.\ \cite{knap12PhysRevX.2.041020} in their investigation
of the Anderson orthogonality catastrophe with the exception that 1D bosons
instead of 3D fermions constitute the majority atoms.

We model this system, which describes potential scattering of bosons in one
dimension, by the grand canonical Hamiltonian in momentum space,
\begin{align} \label{H}
\hat{H}  &  =\sum_{\mathbf{k}}\left(  \epsilon_{\mathbf{k}}-\mu\right)
\hat{b}_{\mathbf{k}}^{\prime\dag}\hat{b}_{\mathbf{k}}^{\prime}+\frac{g_{BB}%
}{2\mathcal{V}}\sum_{\mathbf{k},\mathbf{k}^{\prime}\mathbf{q}}\hat
{b}_{\mathbf{k}+\mathbf{q}}^{\prime\dag}\hat{b}_{\mathbf{k}^{\prime
}-\mathbf{q}}^{\prime\dag}\hat{b}_{\mathbf{k}^{\prime}}^{\prime}\hat
{b}_{\mathbf{k}}^{\prime}
\nonumber\\
&\qquad  +\frac{g_{IB}}{\mathcal{V}}\sum_{\mathbf{k},\mathbf{k}^{\prime}}\hat
{b}_{\mathbf{k}}^{\prime\dag}\hat{b}_{\mathbf{k}^{\prime}}^{\prime},
\end{align}
where $\mathcal{V}$ is the quantization length and $\hat{b}_{\mathbf{k}}^{\prime}$
($\hat{b}_{\mathbf{k}}^{\prime\dag}$) is the field operator for annihilating
(creating) a boson of mass $m_{B}$ with momentum $\mathbf{k}$ and energy
$\epsilon_{\mathbf{k}}=k^{2}/2m_{B}$.  The first line in Eq.\ (\ref{H}) is
the Hamiltonian for background bosons, with $\mu$ the chemical potential and
$g_{BB}\left(  >0\right)  $ the effective 1D\ $s$-wave interaction strength
between two bosons. The second line in Eq.\ (\ref{H}) describes scattering
between bosons and a localized impurity through a delta
function potential with an effective 1D strength $g_{IB}$, which is fixed by
the $s$-wave interaction between the impurity and a background boson.

Instead of $g_{IB}$ and $g_{BB}$, we may also measure two-body $s$-wave
interactions with the corresponding effective 1D scattering lengths $a_{IB}$
and $a_{BB}$. In one dimension, a two-body delta potential with strength $g_{1D}$ can
be shown to produce an $s$-wave scattering amplitude $f\left(  k\right)
=-1/\left(  1+ika_{1D}\right)$, where $a_{1D}=-1/m_{r}g_{1D}$ and is defined as
the 1D $s$-wave scattering length and $m_{r}$ is the reduced mass between two
colliding particles \cite{olshanii98PhysRevLett.81.938}. Thus, for the case
of an infinitely heavy impurity, the two descriptions are related to each other
according to
\begin{equation}
g_{BB}=\frac{-2}{m_{B}a_{BB}},
\qquad
g_{IB}=\frac{-1}{m_{B}a_{IB}}.
\end{equation}
In cold-atom systems, effective 1D interaction strengths (and hence also their
corresponding scattering lengths) can be tuned from negative to positive via
confinement-induced resonance and are related to their 3D counterparts
following well-established recipes, irrespective of whether the impurity and
host bosons experience the same
\cite{olshanii98PhysRevLett.81.938,bergeman03PhysRevLett.91.163201} or
different \cite{peano05NewJournalOfPhysics.7.192} trap frequencies.

As in our earlier publication \cite{kain16PhysRevA.94.013621}, we limit our
study to near zero temperatures where bosons are assumed to be in the deep
BEC regime, i.e.\ there is a macroscopic occupation by bosons
of the condensed ($\mathbf{k}=0$) mode. The Hohenberg-Mermin-Wagner
theorem only prohibits infinite 1D Bose gases from forming a BEC
\cite{hohenberg67PhysRev.158.383,mermin66PhysRevLett.17.1133}, and thus does
not apply to quasi-1D Bose gases in actual cold-atom experiments, which are
neither strictly 1D nor infinite in size. By assuming the
existence of a $\mathbf{k}=0$ BEC mode, we are, in essence, anticipating
future applications of our theory to relatively large trapped gases in one
dimension where true BECs exist at temperatures near zero
\cite{petrov00PhysRevLett.85.3745}. 

In the spirit of the Bogoliubov approximation, we separate out the condensed
mode $\mathbf{k}=0$, treating $\hat{b}_\mathbf{0}'$ as the
classical field ($c$-number) $b_{0}^{\prime}$, and replace the noncondensed
fields $\hat{b}_{\mathbf{k}\neq0}^{\prime}$ in favor of the phonon fields
$\hat{b}_{\mathbf{k}}$ via the Bogoliubov transformation
\begin{equation}
\hat{b}_{\mathbf{k}}^{\prime}=u_{\mathbf{k}}\hat{b}_{\mathbf{k}}%
-v_{\mathbf{k}}\hat{b}_{-\mathbf{k}}^{\dag},\label{phonon field operator}%
\end{equation}
where
\begin{subequations}
\label{uv}%
\begin{align}
u_{\mathbf{k}} &  =\sqrt{\frac{1}{2}\left(  \frac{\epsilon_{\mathbf{k}}%
+g_{BB}n_{B}}{\omega_{\mathbf{k}}}+1\right)  }\label{u}\\
v_{\mathbf{k}} &  =\sqrt{\frac{1}{2}\left(  \frac{\epsilon_{\mathbf{k}}%
+g_{BB}n_{B}}{\omega_{\mathbf{k}}}-1\right)  }\label{v}%
\end{align}
and
\end{subequations}
\begin{equation}
\omega_{\mathbf{k}}=\sqrt{\epsilon_{\mathbf{k}}\left(  \epsilon_{\mathbf{k}%
}+2g_{BB}n_{B}\right)  }=v_{B}k\sqrt{1+\left(  \xi_{B}k\right)  ^{2}%
},\label{omega_k}%
\end{equation}
where $v_{B}=\sqrt{n_{B}g_{BB}/m_{B}}$ is the phonon speed and
\begin{equation}
\xi_{B}=1/\sqrt{4m_{B}n_{B}g_{BB}}\label{healing length}%
\end{equation}
is the healing length.

The Bogoliubov approximation divides the boson-boson interaction---the
term associated with $g_{BB}$ in Eq.\ (\ref{H})---into two pieces that depend on
whether or not the condensed bosons participate in the scattering process.
The piece in which all scattering partners come from noncondensed modes
(and is what Grusdt et al.\
\cite{grusdt17PhysRevA.96.013607,grusdt17NewJournalOfPhysics.19.103035} called
the phonon-phonon interaction) is neglected in the Bogoliubov approach. As
such, within Bogoliubov theory, the Bose gas can be approximated as
consisting of a condensate with number (line) density $n_{B}=\left\vert
b_{0}^{\prime}\right\vert ^{2}/\mathcal{V}$ and chemical potential $\mu
=n_{B}g_{BB}$, and a collection of noninteracting phonons that obey the
dispersion spectrum $\omega_{\mathbf{k}}$ in Eq.\ (\ref{omega_k}).  This
description was shown to hold quite well by Lieb and Liniger (who solved the
1D Bose system exactly and analytically) in the weak-interacting regime, where
the dimensionless coupling strength
\begin{equation}
\gamma\equiv\frac{m_{B}\left\vert g_{BB}\right\vert }{n_{B}}=\frac{2}%
{n_{B}\left\vert a_{BB}\right\vert } \label{gamma}%
\end{equation}
is limited to $\gamma\leq2$ 
\cite{lieb63PhysRev.130.1605,lieb63PhysRev.130.1616}. $\gamma$ in Eq.\
(\ref{gamma}) is defined as the ratio of the interaction energy scale
$n_{B}| g_{BB}|$ to the kinetic energy scale $n_{B}%
^{2}/m_{B}$ so that in one dimension, the lower the boson number density, the stronger the
interaction.  In the strong-interacting regime, where $\gamma>2$, an accurate
description must go beyond the Bogoliubov theory by including
phonon-phonon interactions \cite{,grusdt17NewJournalOfPhysics.19.103035}.  

Having separated out the condensed mode, we replace $\hat{b}_{\mathbf{k}%
}^{\prime}$ with the phonon field operator $\hat{b}_{\mathbf{k}}$ and change
Hamiltonian (\ref{H}) into
\cite{shchadilova16PhysRevLett.117.113002,grusdt17PhysRevA.96.013607,grusdt17NewJournalOfPhysics.19.103035}%
\begin{equation}
\hat{H}=\hat{H}_{1}+\hat{H}_{2},\label{H 1}%
\end{equation}
where%
\begin{align}
\hat{H}_{1} &  =n_{B}g_{IB}+\sum_{\mathbf{k}}\omega_{\mathbf{k}}\hat
{b}_{\mathbf{k}}^{\dag}\hat{b}_{\mathbf{k}}
\nonumber\\
&\qquad  +\frac{1}{\sqrt{\mathcal{V}}}\sum_{\mathbf{k}}g_{\mathbf{k}}\left(  \hat
{b}_{\mathbf{k}}+\hat{b}_{\mathbf{k}}^{\dag}\right)  \label{H 1 1}%
\end{align}
and%
\begin{align}
\hat{H}_{2} &  =\frac{g_{IB}}{\mathcal{V}}\sum_{\mathbf{k}}v_{\mathbf{k}}%
^{2}+\frac{1}{\mathcal{V}}\sum_{\mathbf{kk}^{\prime}}g_{\mathbf{kk}^{\prime}%
}^{+}\hat{b}_{\mathbf{k}}^{\dag}\hat{b}_{\mathbf{k}^{\prime}}
\nonumber\\
&\qquad +  \frac{1}{2\mathcal{V}}\sum_{\mathbf{kk}^{\prime}}g_{\mathbf{kk}^{\prime}%
}^{-}\left(  \hat{b}_{\mathbf{k}}^{\dag}\hat{b}_{\mathbf{k}^{\prime}}^{\dag
}+\hat{b}_{\mathbf{k}}\hat{b}_{\mathbf{k}^{\prime}}\right)  ,\label{H 1 2}%
\end{align}
with $g_{\mathbf{k}}$ and $g_{\mathbf{kk}^{\prime}}^{\pm}$ defined as
\begin{subequations}
\label{g}%
\begin{align}
g_{\mathbf{k}} &  =g_{IB}\sqrt{n_{B}}\chi_{\mathbf{k}}\label{g_k}\\
g_{\mathbf{kk}^{\prime}}^{\pm} &  =\frac{g_{IB}}{2}\left(  \chi_{\mathbf{k}%
}\chi_{\mathbf{k}^{\prime}}\pm\chi_{\mathbf{k}}^{-1}\chi_{\mathbf{k}^{\prime}%
}^{-1}\right),\label{g_kk'}%
\end{align}
where%
\end{subequations}
\begin{equation}
\chi_{\mathbf{k}}=\sqrt{\epsilon_{\mathbf{k}}/\omega_{\mathbf{k}}}.
\end{equation}
$\hat{H}_{1}$ in Eq.\ (\ref{H 1 1}) represents the usual Fr\"{o}hlich
Hamiltonian in the heavy impurity limit.  The second line in Eq.
(\ref{H 1 1}), which is traced to impurity scattering of a boson from the
condensed mode to a noncondensed mode, now represents single-phonon
scattering.  $\hat{H}_{2}$ in Eq.\ (\ref{H 1 2}) represents the part that goes
beyond the Fr\"{o}hlich paradigm. The term $g_{IB} \sum_{\mathbf{k}}v_{\mathbf{k}}^{2}/\mathcal{V}$ arises from normal ordering.  The remaining
terms in Eq.\ (\ref{H 1 2}) describe two-phonon scattering, which is
traced to impurity scattering of a boson between two noncondensed modes and is
therefore important in the limit of strong impurity-boson interactions.

\section{Mean-Field Solution}

In the absence of $H_{2}$, $\hat{H}=\hat{H}_{1}$ in Eq.\
(\ref{H 1 1}), which has the same mathematical form as the electron-phonon
Hamiltonian when the electrons are localized in space.
This Hamiltonian is known to have the exact solution
\cite{Mahan00Book}
\begin{equation}
\left\vert z\right\rangle =
{\displaystyle\prod\limits_{\mathbf{k}}}
\left\vert z_{\mathbf{k}}\right\rangle ,\label{|z>}%
\end{equation}
where
\begin{equation}
\left\vert z_{\mathbf{k}}\right\rangle =\exp\left(  z_{\mathbf{k}}\hat
{b}_{\mathbf{k}}^{\dag}-z_{\mathbf{k}}^{\ast}\hat{b}_{\mathbf{k}}\right)
\left\vert 0\right\rangle
\end{equation}
is the coherent state of mode \textbf{k} and $z_{\mathbf{k}}=-g_{\mathbf{k}%
}/(  \omega_{\mathbf{k}}\sqrt{\mathcal{V}})$. 
The mean-field variational approach
\cite{lee53PhysRev.90.297,shashi14PhysRevA.89.053617,shchadilova16PhysRevLett.117.113002}
amounts to assuming that even when $H_{2}$ is included, the
ground state continues to be in the product state (\ref{|z>}), but with
$z_{\mathbf{k}}$ a variational parameter to be determined.
The expectation value of the Hamiltonian in the coherent state is
\begin{equation}
\mathcal{E}\left(  z,z^{\ast}\right)  =\left\langle z\right\vert \hat
{H}(  \hat{b},\hat{b}^{\dag})  \left\vert z\right\rangle =\hat
{H}(  z,z^{\ast}),
\end{equation}
where in the second equality we used that 
$\hat{H}(  \hat{b},\hat{b}^{\dag})$ in Eqs.\ (\ref{H 1 1}) and
(\ref{H 1 2})
is in a normally ordered form.  Minimizing the energy $\mathcal{E}(z,z^*)$
with respect to $z_{\mathbf{k}}$ 
leads to the saddle point equation
\begin{equation}
\omega_{\mathbf{k}}z_{\mathbf{k}}+\frac{\chi_{\mathbf{k}}}{\mathcal{V}}%
g_{IB}\sum_{\mathbf{k}^{\prime}}\chi_{\mathbf{k}^{\prime}}z_{\mathbf{k}%
^{\prime}}=-\frac{g_{\mathbf{k}}}{\sqrt{\mathcal{V}}}%
.\label{saddle point condition}%
\end{equation}
\newline In arriving at Eq.\ (\ref{saddle point condition}), we have taken
$z_{\mathbf{k}}$ to be purely real because the coupling constant
$g_{\mathbf{k}}$ is a real number.  The mean-field polaron energy, defined as
$\mathcal{E}_{0}\equiv\mathcal{E}\left(  z,z^{\ast}\right)  $ at the saddle
point, simplifies to%

\begin{equation}
\mathcal{E}_{0}=n_{B}g_{IB}+\frac{1}{\sqrt{\mathcal{V}}}\sum_{\mathbf{k}%
}g_{\mathbf{k}}z_{\mathbf{k}}+\frac{g_{IB}}{\mathcal{V}}\sum_{\mathbf{k}%
}v_{\mathbf{k}}^{2}.\label{E_0 saddle}%
\end{equation}
When $z_{\mathbf{k}}$ in Eq.\ (\ref{E_0 saddle}) is replaced with the solution to
Eq.\ (\ref{saddle point condition}),
\begin{equation}
z_{\mathbf{k}}=-\frac{1}{\sqrt{\mathcal{V}}}\frac{g_{\mathbf{k}}%
/\omega_{\mathbf{k}}}{1+\frac{g_{IB}}{\mathcal{V}}\sum_{\mathbf{k}}\frac
{\chi_{\mathbf{k}}^{2}}{\omega_{\mathbf{k}}}},\label{z_k}%
\end{equation}
the mean-field polaron energy takes its final form,
\begin{equation}
\mathcal{E}_{0}=\mathcal{E}_{0}^{M}+\frac{g_{IB}}{\mathcal{V}}\sum
_{\mathbf{k}}v_{\mathbf{k}}^{2},\label{E_0 mean-field}%
\end{equation}
where%
\begin{equation}
\mathcal{E}_{0}^{M}\equiv\frac{n_{B}}{\frac{1}{g_{IB}}+\frac{1}{\mathcal{V}%
}\sum_{\mathbf{k}}\frac{\chi_{\mathbf{k}}^{2}}{\omega_{\mathbf{k}}}%
}.\label{E mean-field}%
\end{equation}

The last term, $g_{IB}\sum_{\mathbf{k}}v_{\mathbf{k}}^{2}/\mathcal{V}$, arises
from the normal ordering of $\hat{H}_{2}$ in Eq.\ (\ref{H 1 2}) and thus represents
vacuum energy contributed by quantum fluctuations associated with the
interaction between the impurity and noncondensed bosons.  In the infrared
limit, where the low-momentum cutoff, $\lambda$, approaches zero, the term $g_{IB}\sum_{\mathbf{k}}v_{\mathbf{k}}^{2}/\mathcal{V}$ diverges
with $\lambda$ logarithmically as $-g_{IB}\sqrt{m_{B}n_{B}g_{BB}}\left(
\ln\lambda\right)  /2\pi$.  This log-divergence can be traced to the 1D
density of states being inversely proportional to the square root of the
energy, leading to a dramatic enhancement of quantum fluctuations in the
IR limit. This same enhancement was at the heart of  the
Hohenberg-Mermin-Wagner theorem
\cite{hohenberg67PhysRev.158.383,mermin66PhysRevLett.17.1133}, which precludes
a true BEC from forming in a true 1D infinite Bose system. Following
Grusdt et al.\ \cite{grusdt17NewJournalOfPhysics.19.103035}, we remove the
divergent term from $\mathcal{E}_{0}$, treating $\mathcal{E}_{0}^{M}$ as ``the
polaron energy in the mean-field theory."  The quotation marks are meant to
indicate that $\mathcal{E}_{0}^{M}$, obtained in such a brute-force manner,
should not be regarded as the result of a consistent theory.  That it agrees
well with the exact result, which is presented in the next section, means only that $\mathcal{E}_{0}^{M}$ may serve as a good measuring stick for our exact
theory.


\section{Exact Solution}

In this section, we solve $\hat{H}$ exactly and obtain a polaron energy free of
the IR divergence.  We start by replacing the phonon field operators $\hat
{b}_{\mathbf{k}}$ and $\hat{b}_{\mathbf{k}}^{\dag}$ with the shifted phonon field
operators
\begin{subequations}
\label{shifted field operator}%
\begin{align}
\hat{c}_{\mathbf{k}}  &  =\hat{b}_{\mathbf{k}}-z_{\mathbf{k}}\\
\hat{c}_{\mathbf{k}}^{\dag}  &  =\hat{b}_{\mathbf{k}}^{\dag}-z_{\mathbf{k}%
}^{\ast},
\end{align}
which describe quantum fluctuations around $z_{\mathbf{k}}$, the saddle-point
solution in Eq.\ (\ref{z_k}). By virtue of the saddle-point condition in Eq.
(\ref{saddle point condition}), the Hamiltonian in terms of the shifted
operators is free of the Fr\"{o}hlich terms (those linear in $\hat
{c}_{\mathbf{k}}$ and $\hat{c}_{\mathbf{k}}^{\dag}$) and is given by%

\end{subequations}
\begin{align}
\hat{H} &  =\mathcal{E}_{0}+\sum_{\mathbf{k}}\omega_{\mathbf{k}}\hat
{c}_{\mathbf{k}}^{\dag}\hat{c}_{\mathbf{k}}+\frac{1}{\mathcal{V}}%
\sum_{\mathbf{kk}^{\prime}}g_{\mathbf{kk}^{\prime}}^{+}\hat{c}_{\mathbf{k}%
}^{\dag}\hat{c}_{\mathbf{k}^{\prime}}
\nonumber\\
&\qquad  +\frac{1}{2\mathcal{V}}\sum_{\mathbf{kk}^{\prime}}g_{\mathbf{kk}^{\prime}%
}^{-}\left(  \hat{c}_{\mathbf{k}}^{\dag}\hat{c}_{\mathbf{k}^{\prime}}^{\dag
}+\hat{c}_{\mathbf{k}}\hat{c}_{\mathbf{k}^{\prime}}\right)
,\label{H quadratic}%
\end{align}
which is quadratic and is therefore exactly solvable. 

Following the standard procedure, we introduce the quasiparticle field
operator $\hat{d}_{n}$ through the generalized Bogoliubov transformation
\begin{subequations}
\label{Bogoliubov transformation}%
\begin{align}
\hat{d}_{n}  &  =\sum_{\mathbf{k}}\left(  U_{n\mathbf{k}}^{\ast}\hat
{c}_{\mathbf{k}}-V_{n\mathbf{k}}^{\ast}\hat{c}_{\mathbf{k}}^{\dag}\right)  \\
\hat{d}_{n}^{\dag}  &  =\sum_{\mathbf{k}}\left(  U_{n\mathbf{k}}\hat
{c}_{\mathbf{k}}^{\dag}-V_{n\mathbf{k}}\hat{c}_{\mathbf{k}}\right)  ,
\end{align}
where $U$ and $V$ are $\mathcal{N}\times\mathcal{N}$ matrices, with
$\mathcal{N}$ being the total number of modes in momentum $\mathbf{k}$ space.
The $n$th row of the $U$ and $V$ matrices contains the $n$th eigenstate,
$\left(  U_{n},V_{n}\right)  ^{T}$, of the eigenvalue equation
\end{subequations}
\begin{equation}
M\left(
\begin{array}
[c]{c}%
U_{n}\\
V_{n}%
\end{array}
\right)  =w_{n}\left(
\begin{array}
[c]{c}%
U_{n}\\
V_{n}%
\end{array}
\right)  , \label{M}%
\end{equation}
where $M$ is the $2\mathcal{N}\times2\mathcal{N}$ matrix
\begin{equation}
M\equiv\left(
\begin{array}
[c]{cc}%
A & B\\
-B & -A
\end{array}
\right),
\end{equation}
with $A$ and $B$ being the $\mathcal{N}\times\mathcal{N}$ matrices, whose
components are
\begin{align}
A_{\mathbf{kk}^{\prime}}  &  =\omega_{\mathbf{k}}\delta_{\mathbf{k}%
,\mathbf{k}^{\prime}}+\frac{g_{\mathbf{kk}^{\prime}}^{+}}{\mathcal{V}}\\
B_{\mathbf{kk}^{\prime}}  &  =\frac{g_{\mathbf{kk}^{\prime}}^{-}}{\mathcal{V}%
}.
\end{align}

Note that the generalized Bogoliubov transformation (\ref{Bogoliubov
transformation}) mixes annihilation and creation operators in exactly the same
manner as the multimode squeezing operator in quantum optics
\cite{walls08Book}.
In the beyond-Fr\"{o}hlich model, squeezing is traced to simultaneous creation or annihilation of two noncondensed bosons by impurity scattering, which are nonlinear matter wave mixing processes akin to parametric up and down conversion of light waves, which are responsible for the squeezing phenomenon in quantum optics.

If all eigenvalues are real, we can cast Hamiltonian (\ref{H quadratic})
into the diagonal form%
\begin{align}
\hat{H} &  =\mathcal{E}_{0}^{M}+\sum_{n}w_{n}\hat{d}_{n}^{\dag}\hat{d}%
_{n}+\frac{1}{2}\left(  \sum_{n}w_{n}-\sum_{\mathbf{k}}\omega_{\mathbf{k}%
}\right)  
\nonumber\\
&\qquad  +\frac{g_{IB}}{\mathcal{V}}\sum_{\mathbf{k}}v_{\mathbf{k}}^{2}-\frac{1}%
{2}\sum_{\mathbf{k}}\frac{g_{\mathbf{kk}}^{+}}{\mathcal{V}}\label{H d_n},
\end{align}
which is
in terms of $\hat{d}_{n}$, where the sum over $n$ is only over eigenstates
$\left\vert w_{n}\right\rangle $ with positive norm $\left\langle
w_{n}\right\vert \zeta\left\vert w_{n}\right\rangle >0$, where $\zeta$ is the
$2\mathcal{N}\times2\mathcal{N}$ matrix%
\begin{equation}
\zeta=\left(
\begin{array}
[c]{cc}%
I & 0\\
0 & -I
\end{array}
\right)  ,
\end{equation}
with $I$ the $\mathcal{N}\times\mathcal{N}$ identity matrix.  In other words,
the sum over $n$ includes\ only those eigenmodes which can be normalized to +1
with metric $\zeta$:
\begin{equation}
\left\langle w_{n}\right\vert \zeta\left\vert w_{n}\right\rangle =1,
\end{equation}
or explicitly
\begin{equation}
\sum_{\mathbf{k}}\left(  \left\vert U_{n\mathbf{k}}\right\vert ^{2}-\left\vert
V_{n\mathbf{k}}\right\vert ^{2}\right)  =1.\label{U^2 - V^2}%
\end{equation}
The fact that $\hat{H}$ in Eq.\ (\ref{H quadratic}) is exactly solvable by the
generalized Bogoliubov transformation means that the ground polaron state for
the beyond-Fr\"{o}hlich model is an exact multimode squeezing state.

Equation (\ref{M}) is the bosonic analog of the Bogoliubov--de Gennes (BDG)
equation for fermions in the state of superfluid pairings. In contrast to the
matrix in the BDG equation for fermions, which is Hermitian, the matrix $M$ in
the BDG equation for bosons [Eq.\ (\ref{M})] is non-Hermitian and its
eigenvalues can be complex. Just as there exists an intrinsic anti-unitary
particle-hole symmetry in the fermionic BDG \cite{hasan10RevModPhys.82.3045}, there exists, in the bosonic BDG equation
(\ref{M}), an analogous built-in symmetry:
\begin{equation}
\tau M\tau^{-1}=-M,
\label{particle-hole}%
\end{equation}
which maps $M$ to $-M$, with $\tau$ being the orthogonal matrix
\begin{equation}
\tau=\left(
\begin{array}
[c]{cc}%
0 & I\\
I & 0
\end{array}
\right)  .
\end{equation}
As a consequence of this ``particle-hole" symmetry, for every eigenvector
$\left\vert w_{n}\right\rangle =\left(  U_{n},V_{n}\right)  ^{T}$ with a
nonvanishing eigenvalue $w_{n}$, there corresponds an eigenvector
\begin{equation}
\left\vert -w_{n}\right\rangle =\tau\left\vert w_{n}\right\rangle =\left(
V_{n},U_{n}\right)  ^{T}%
\end{equation}
with an eigenvalue $-w_{n}$. 
Eigenvalues in our system therefore appear in pairs with opposite signs

Thus, there arise three possible scenarios for stability of the system. (a)
The system is said to be \textit{dynamically unstable} if one or more pairs of
eigenvalues are complex---an exponentially small perturbation can cause the
system to depart irreversibly from equilibrium. In the absence of any pairs
of complex eigenvalues, (b) the system is said to be \textit{thermodynamically
unstable} if one or more eigenvalues associated with eigenvectors normalizable
to $+1$ with metric $\zeta$ are negative---such a state cannot be created
adiabatically by gradually reducing the entropy associated with the thermal
energy---and (c) the system is said to be \textit{thermodynamically stable} if all
eigenvalues associated with eigenvectors normalized to $+1$ with metric $\zeta$
are positive.

In cases (b) and (c), i.e.\ those without complex eigenvalues, Eq.\
(\ref{H d_n}) holds true.  We are thus led to define
\begin{equation}
E_{0}=\mathcal{E}_{0}^{M}+\frac{1}{2}\left(  \sum_{n}w_{n}-\sum_{\mathbf{k}%
}\omega_{\mathbf{k}}\right)  -\frac{g_{IB}}{2\mathcal{V}}\sum_{\mathbf{k}%
}\label{E_0}%
\end{equation}
as the polaron energy in a metastable state for case (b) and the polaron
energy in the ground state for case (c). 

In case (a), Eq.\ (\ref{H d_n}) does not apply because modes with
complex eigenvalues arise. 
Such modes always have a vanishing norm with respect to
metric $\zeta$ and 
are therefore excluded from Eq.\ (\ref{E_0}), where
the sum is limited only to modes normalizable to $+1$ with metric $\zeta$.
Thus, 
for case (a), Eq.\ (\ref{E_0}), in fact, is well defined. In the present
study, we continue to use Eq.\ (\ref{E_0}) as the ``polaron energy" for case
(a).  
For a polaron system where complex eigenvalues are all purely imaginary, Eq.\ (\ref{E_0}) gives the exact polaron energy in the limit of vanishingly small imaginary eigenvalues.
In this situation, for all practical purposes, the polaron can be considered as dynamically stable.
Since in our system, complex eigenvalues are purely imaginary and have small imaginary parts (as we discuss in the next section), the use of Eq.\ (\ref{E_0}) for the polaron energy is not particularly unreasonable.
  
A comment is in order concerning IR and ultraviolet (UV) divergences. The last term in Eq.\
(\ref{H d_n}), when making use of $g_{\mathbf{kk}}^{+}$ in Eq.\ (\ref{g_kk'})
and $v_{\mathbf{k}}$ in Eq.\ (\ref{v}), becomes%
\begin{equation}
\frac{1}{2}\sum_{\mathbf{k}}\frac{g_{\mathbf{kk}}^{+}}{\mathcal{V}}%
=\frac{g_{IB}}{\mathcal{V}}\sum_{\mathbf{k}}v_{\mathbf{k}}^{2}+\frac{g_{IB}%
}{2\mathcal{V}}\sum_{\mathbf{k}},
\end{equation}
\newline which is found to contain an identical IR divergence term $g_{IB}%
\sum_{\mathbf{k}}v_{\mathbf{k}}^{2}/\mathcal{V}$.  This explains how the sum
of the last two terms in Eq.\ (\ref{H d_n}) eliminates the IR divergence, but
it gives rise to a UV divergence represented by the last term in Eq.\ (\ref{E_0}).
 The middle term in Eq.\ (\ref{E_0}) (which in a Fermi polaron system
\cite{kain17PhysRevA.96.033627} is finite because of the Fermi surface) is
found (numerically) to contain a UV divergence that is identical and therefore
cancels the UV divergent term in Eq.\ (\ref{E_0}).  In conclusion, the polaron
energy in one dimension, when quantum fluctuations are taken care of exactly,
are free of both IR and UV divergences.

Figure \ref{Fig:polaronEnergy} compares, within the weakly (Bose-Bose)
interacting regime, the exact polaron energy $E_{0}$ (solid black curves), given by
Eq.\ (\ref{E_0}), with the mean-field polaron energy $\mathcal{E}_{0}^{M}$
(dashed blue curves), given by Eq.\ (\ref{E mean-field}). The energies are given
as a function of $\eta$, where $\eta$ is the dimensionless parameter 
\begin{equation}
\eta=\frac{g_{IB}}{g_{BB}}, \label{eta}%
\end{equation}
which measures the impurity-boson interaction relative to the boson-boson
interaction.  For $\eta>0$, the polaron is repulsive and the energy increases
with $\eta$ monotonically from 0 until it saturates in the limit of large
positive $\eta$.  For $\eta<0$, the polaron changes from attractive to
repulsive as $\eta$ decreases across a critical value,
\begin{equation}
\eta_{c}=-2/\sqrt{\gamma},
\end{equation}
at which the denominator in $\mathcal{E}_{0}^{M}$ [Eq.\ (\ref{E mean-field})]
vanishes, the details of which we present in the next section.  When the
polaron energy is plotted as a function of $\eta^{-1}$, it becomes evident
that the two branches of repulsive polarons, which are separated in $\eta$
space, are actually adiabatically connected---there is a smooth crossover
when $\eta^{-1}$ changes across $0$ (unitarity).
\begin{figure}
\centering
\includegraphics[
width=3.4in
]%
{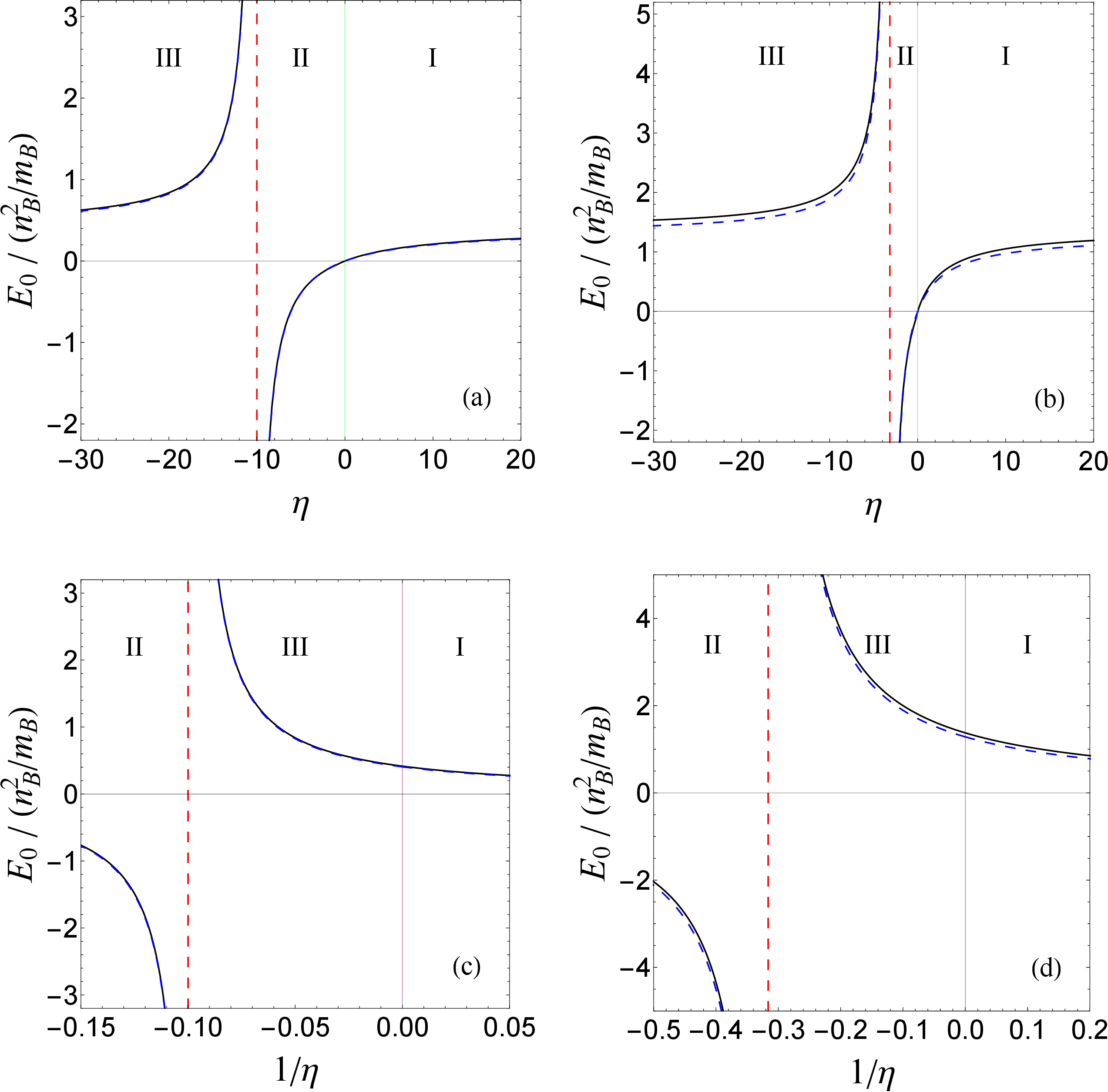}
\caption{The top row displays the polaron energy as a function of $\eta = g_{IB}/g_{BB}$ and the bottom row as a function of $1/\eta$.  The exact polaron energy, $E_0$, is represented by solid black curves and the mean field polaron energy, $\mathcal{E}_0^M$, by dashed blue curves.  The left column, (a) and (c), has $\gamma = 0.04$ and the right column, (b) and (d), has $\gamma = 0.4$.  The dashed red line marks the critical value $\eta_c = -2/\sqrt{\gamma}$.  In each plot I, II, and III indicate phases, which are explained in Sec.\ \ref{sec:eigenvalues} and in the caption to Fig.\ \ref{Fig:phaseDiagram}.
}
\label{Fig:polaronEnergy}%
\end{figure}

It can be seen that quantum fluctuations, which are absent in the mean-field
approach, offset almost entirely the IR-divergent term in $\mathcal{E}_{0}$,
the last term in Eq.\ (\ref{E_0 mean-field}), so that the exact polaron energy
$E_{0}$ remains fairly close to $\mathcal{E}_{0}^{M}$.  That the two
results are very close to each other, even in the limit of large $\left\vert
\eta\right\vert $, is expected to be the case only in the limit of an
infinitely heavy impurity.  For a mobile impurity, impurity recoil
contributes additional terms to the Hamiltonian that are inversely
proportional to the impurity mass.  As the impurity becomes lighter, $E_{0}$
is expected to become increasingly different from $\mathcal{E}_{0}^{M}$,
especially in the strong-coupling regime where $\left\vert \eta\right\vert $
is large \cite{grusdt17NewJournalOfPhysics.19.103035}.  


\section{Eigenvalue Analysis and Exact Phase Diagrams}
\label{sec:eigenvalues}

We can gain significant insight into the stability of the system by analyzing
the eigenvalue matrix equation. We now work to solve the eigenvalue and
eigenvector problem from the coupled equations. (In this section, when no
confusion is likely to arise, we drop the subscript $n$ and write
$w_{n},U_{n\mathbf{k}}$, and $V_{n\mathbf{k}}$ as $w,U_{\mathbf{k}}$, and
$V_{\mathbf{k}}$ for notational simplicity.) We first define, for each
$\mathbf{k}$, the two variables $X_{\mathbf{k}}^{\pm}$ as
\begin{equation}
\left(
\begin{array}
[c]{c}%
X_{\mathbf{k}}^{+}\\
X_{\mathbf{k}}^{-}%
\end{array}
\right)  =\frac{1}{2}\left(
\begin{array}
[c]{cc}%
1 & 1\\
1 & -1
\end{array}
\right)  \left(
\begin{array}
[c]{c}%
U_{\mathbf{k}}\\
V_{\mathbf{k}}%
\end{array}
\right)  
\end{equation}
and transform Eq.\ (\ref{M}) into
\begin{equation}
\left(
\begin{array}
[c]{c}%
X_{\mathbf{k}}^{+}\\
X_{\mathbf{k}}^{-}%
\end{array}
\right)  =\frac{g_{IB}}{\mathcal{V}}\left(
\begin{array}
[c]{c}%
\frac{\chi_{\mathbf{k}}\omega_{\mathbf{k}}}{w^{2}-\omega_{\mathbf{k}}^{2}%
},\frac{\chi_{\mathbf{k}}^{-1}w}{w^{2}-\omega_{\mathbf{k}}^{2}}\\
\frac{\chi_{\mathbf{k}}w}{w^{2}-\omega_{\mathbf{k}}^{2}},\frac{\chi
_{\mathbf{k}}^{-1}\omega_{\mathbf{k}}}{w^{2}-\omega_{\mathbf{k}}^{2}}%
\end{array}
\right)  \left(
\begin{array}
[c]{c}%
a^{+}\\
a^{-}%
\end{array}
\right)  ,
\end{equation}
where $a^{\pm}$ are constants defined as
\begin{equation}
a^{+}=\sum_{\mathbf{k}}\chi_{\mathbf{k}}X_{\mathbf{k}}^{+},
\qquad
a^{-}
=\sum_{\mathbf{k}}\chi_{\mathbf{k}}^{-1}X_{\mathbf{k}}^{-}.
\end{equation}
This leads to the following linearly coupled homogeneous equations for
$a^{\pm}$:
\begin{equation}
\left(
\begin{array}
[c]{cc}%
g_{IB}^{-1}+I_{+}\left(  w^{2}\right)   & wI_{0}\left(  w^{2}\right)  \\
wI_{0}\left(  w^{2}\right)   & g_{IB}^{-1}+I_{-}\left(  w^{2}\right)
\end{array}
\right)  \left(
\begin{array}
[c]{c}%
a^{+}\\
a^{-}%
\end{array}
\right)  =0,
\end{equation}
where
\begin{subequations}
\label{discrete sum}%
\begin{align}
I_{0}\left(  x\right)   &  =\frac{1}{\mathcal{V}}\sum_{\mathbf{k}}\frac
{1}{\omega_{\mathbf{k}}^{2}-x},\label{I_0}\\
I_{\pm}\left(  x\right)   &  =\frac{1}{\mathcal{V}}\sum_{\mathbf{k}}\frac
{\chi_{\mathbf{k}}^{\pm2}\omega_{\mathbf{k}}}{\omega_{\mathbf{k}}^{2}%
-x}.\label{I_+-}%
\end{align}
The eigenvalues follow from the vanishing of the determinant, and thus are the
roots of the equation
\end{subequations}
\begin{align}
F\left(  w^{2}\right)  &\equiv-w^{2}\left[  I_{0}\left(  w^{2}\right)
\right]  ^{2}
\nonumber\\
&\qquad +\left[  g_{IB}^{-1}+I_{+}\left(  w^{2}\right)  \right]  \left[  g_{IB}%
^{-1}+I_{-}\left(  w^{2}\right)  \right].
\label{F(w^2)=0}
\end{align}
\newline The corresponding eigenvectors are given by
\begin{subequations}
\label{X+X-}%
\begin{align}
U_{\mathbf{k}} &  =\frac{\chi_{\mathbf{k}}}{w-\omega_{\mathbf{k}}}\frac
{f}{\sqrt{\mathcal{V}}}\left[  1-\frac{g_{IB}^{-1}+I_{+}\left(  w^{2}\right)
}{w\chi_{\mathbf{k}}^{2}I_{0}\left(  w^{2}\right)  }\right]  \\
V_{\mathbf{k}} &  =\frac{-\chi_{\mathbf{k}}}{w+\omega_{\mathbf{k}}}\frac
{f}{\sqrt{\mathcal{V}}}\left[  1+\frac{g_{IB}^{-1}+I_{+}\left(  w^{2}\right)
}{w\chi_{\mathbf{k}}^{2}I_{0}\left(  w^{2}\right)  }\right]  ,
\end{align}
where $f$ is determined by the normalization condition (\ref{U^2 - V^2}).%

The system is dynamically unstable if $w$ is a complex number.  For our
system, when complex eigenvalues occur, numerical simulations suggest that there exists no more than a single pair. Then, the complex eigenvalues must be purely imaginary, as we
explain.  In addition to the ``particle-hole" symmetry (\ref{particle-hole}),
our BDG equation has the time-reversal symmetry
\end{subequations}
\begin{equation}
KMK^{-1}=M,
\end{equation}
where $K$ is the complex conjugate operation.  This implies that for every
eigenvector $\left\vert w_{n}\right\rangle =(  U_{n}%
,V_{n})^{T}$ with nonvanishing eigenvalue $w_{n}$, there exists an
eigenvector
\begin{equation}
\left\vert w_{n}^{\ast}\right\rangle 
=K\left\vert w_{n}\right\rangle
=\left(  U_{n}^{\ast},V_{n}^{\ast}\right)^{T}
\end{equation}
with eigenvalue $w_n^*$.  In consequence, for a complex eigenvalue $w_{n}$, in addition to the pair
$w_{n}$ and $-w_{n}$ guaranteed by the ``particle-hole" symmetry, there exists
another pair, $w_{n}^{\ast}$ and $-w_{n}^{\ast}$, guaranteed by the
time-reversal symmetry.  
For real eigenvalues the second pair is redundant, since it is identical to the first pair.  For complex eigenvalues, however, only when they are purely imaginary are the two pairs equivalent.
Thus, in our case where among all pairs of
eigenvalues, only one pair is complex, complex eigenvalues must be purely imaginary.

As a result, our system makes a transition from thermodynamically stable or metastable to
dynamically unstable when $w^{2}$ changes from positive to negative, which
obviously occurs at $w^{2}=0$.  Since Eq.\ (\ref{F(w^2)=0}) must be satisfied,
we have for $w^{2}=0$ either%
\begin{equation}
\frac{1}{g_{IB}}+\frac{1}{\mathcal{V}}\sum_{\mathbf{k}}\frac{\chi_{\mathbf{k}%
}^{-2}}{\omega_{\mathbf{k}}}=0\label{second}%
\end{equation}
or%
\begin{equation}
\frac{1}{g_{IB}}+\frac{1}{\mathcal{V}}\sum_{\mathbf{k}}\frac{\chi_{\mathbf{k}%
}^{2}}{\omega_{\mathbf{k}}}=0.\label{first}%
\end{equation}
The first possibility (\ref{second}) can be shown to be satisfied when
\begin{equation}
g_{IB}=0, \label{critical condition 1}
\end{equation}
(more precisely, $g_{IB}=0^{-}$), which is expected to be unique to one dimension, since
in arriving at it, we made explicit use of the IR divergence---the sum in
Eq.\ (\ref{second}) equals $m_{B}/\pi\lambda$ in the continuous limit and hence
diverges in the IR limit when $\lambda\rightarrow0^{+}$. 
The second condition (\ref{first}) coincides with the mean-field singularity,
where the polaron changes from attractive to repulsive (as mentioned in the
previous section), and is equivalent to
\begin{equation}
g_{IB}=g_{c}\equiv-\sqrt{\frac{4n_{B}g_{BB}}{m_{B}}}.
\label{critical condition 2}%
\end{equation}

Since$\ g_{IB}$ in one dimension has dimensions of energy $\times$ length, we define
\begin{equation}
\bar{g}_{IB}=\frac{g_{IB}}{1/(m_{B}\xi_{B})}%
\end{equation}
as a unitless parameter that measures the impurity-boson $s$-wave interaction in
units of $1/m_{B}\xi_{B}$, where $\xi_{B}$ is the healing length defined in
Eq.\ (\ref{healing length}). The second critical condition
(\ref{critical condition 2}) is then simply
\begin{equation}
\bar{g}_{IB}=\bar{g}_{c}\equiv-1.
\end{equation}
Thus, the $\bar{g}_{IB}$-phase diagram is very simple. It is divided into
phase I, where $\bar{g}_{IB}>0$; phase II, where $-1<\bar{g}_{IB}<0$; and
phase III, where $\bar{g}_{IB}<-1$.  Further, $\bar{g}_{IB}$
is related to $\gamma$ (${>0}$) in Eq.\ (\ref{gamma}) and $\eta$ in
Eq.\ (\ref{eta}) according to
\begin{equation}
\bar{g}_{IB}=\eta\sqrt{\gamma}/2.
\end{equation}
As a result,
for the $(\eta,\gamma)$-phase diagram,
the three segments in the $\bar{g}_{IB}$-phase diagram 
morph into three areas bordered by $\eta\sqrt{\gamma}=-2, \eta=0$, and
$\gamma=0$.
The $\bar{g}_{IB}$- and $(\eta,\gamma)$-phase diagrams are shown in Fig.\ \ref{Fig:phaseDiagram}.
\begin{figure}
\centering
\includegraphics[
width=2.3in
]%
{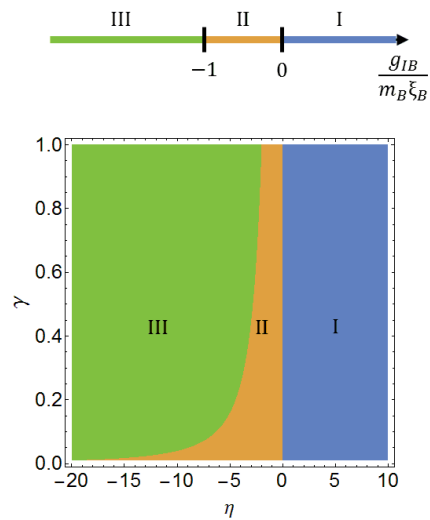}%
\caption{
The three phases of our system:\ thermodynamically stable (phase I), dynamically unstable (phase II), and thermodynamically unstable (phase III).  The phase diagram at the top indicates where these phases occur in the one-dimensional $g_{IB}$-parameter space.  The bottom phase diagram equivalently indicates where these phases occur in the two-dimensional $(\eta,\gamma)$-parameter space.}
\label{Fig:phaseDiagram}%
\end{figure}

We now examine $F(  w^{2})$ in Eq.\ (\ref{F(w^2)=0}) and solve for
its roots, i.e.\ the solutions to $F(  w^{2})  =0$, which allows us to determine the
stability properties of the three phases in the phase diagrams. Figure
\ref{Fig:F(w^2)} plots $F(  w^{2})  $ as a function of $w^{2}$.
 We find that the function contains poles that are regularly spaced at the
locations of $\omega_{\mathbf{k}}^{2}$ with a single root trapped between
adjacent poles. We refer to these as ``regular" roots. The totality
of the regular roots when converted to energy forms the continuum part of the
eigenenergy spectrum. Figure \ref{Fig:F(w^2)}(a) illustrates a typical $F(
w^{2})  $ in phase I (of the phase diagrams), in which every root is a
regular one sandwiched between adjacent poles. 
Figures \ref{Fig:F(w^2)}(c) and \ref{Fig:F(w^2)}(e) illustrate a typical $F(
w^{2})  $ in phases II and III, respectively. 
In phase II, in addition to
the regular roots, there emerges an isolated root that lies at $w^{2}<0$, but
close to the origin. In phase III, $F(  w^{2})  $ again
maintains the same pattern with respect to the regular roots, but now also
contains one additional $w^{2}>0$ root, where two
roots lie between adjacent poles. In the continuum $\mathbf{k}$ limit, one
of the two roots joins the continuum while the other one becomes part of the
discrete spectrum.
\begin{figure}
\centering
\includegraphics[
width=3.3in
]%
{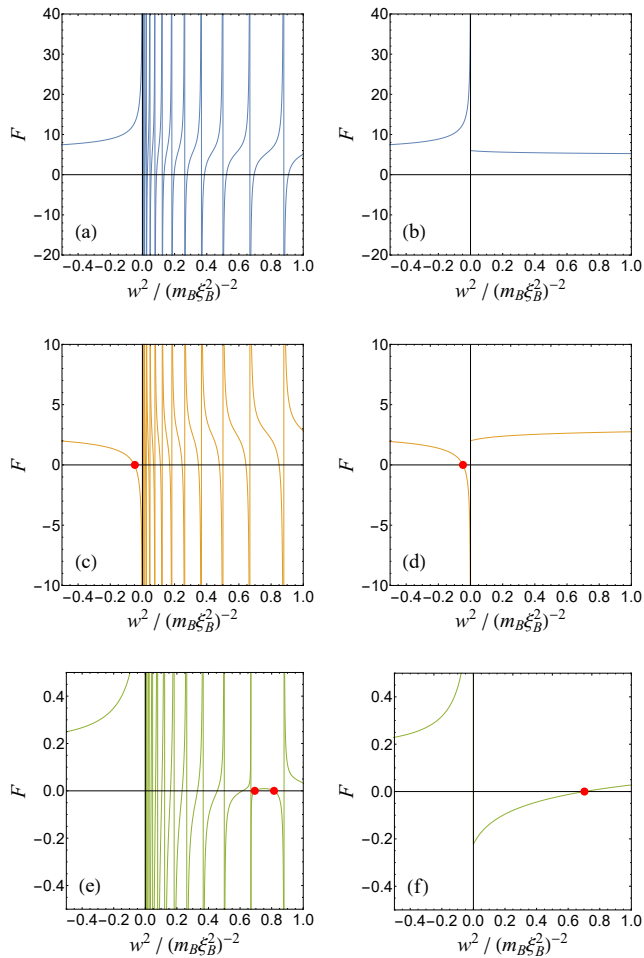}%
\caption{Each row displays the function $F(w^2)$, given in Eq.\ (\ref{F(w^2)=0}), as a function of $w^2$ for one of the three phases.  The roots of $F(w^{2})$, i.e.\ the points at which it crosses the horizontal axis, are the
square of the eigenenergies, $w_{n}^{2}$, of the eigenvalue equation (\ref{M}).  In all plots $\gamma = 0.04$.  The left column displays $F(w^2)$ as computed with the
discrete sums in Eqs.\ (\ref{discrete sum}) and the right column displays $F(w^2)$ in the continuous limit where the
discrete sums are replaced with integrals with respect to momentum.  By comparing the left column with the right, we can see how discrete roots survive in the continuous limit.  The top row, (a) and (b), displays phase I with $\bar{g}_{IB} = g_{IB}/(m_B\xi_B)^{-1}  = 0.5$ (equivalently $\eta = 5$), the middle row, (c) and (d), displays phase II with $\bar{g}_{IB} = -0.5$ ($\eta = -5$), and the bottom row, (e) and (f), displays phase III with $\bar{g}_{IB} = -1.5$ ($\eta = -15$).
}
\label{Fig:F(w^2)}%
\end{figure}

We can determine the discrete spectrum by moving to the continuous
$\mathbf{k}$ limit by converting discrete sums over $\mathbf{k}$ in Eq.\
(\ref{discrete sum}) to integrals over $\mathbf{k}$, which can be evaluated
analytically. We arrive at the results
\begin{equation}
I_{n=0,\pm}\left(  x\right)  =\left\{
\begin{array}
[c]{c}%
I_{n}^{>}\left(  x\right)  \text{, if }x>0\\
I_{n}^{<}\left(  x\right)  \text{, if }x<0
\end{array}
\right.  \text{, }%
\end{equation}
where
\begin{subequations}
\label{I>}%
\begin{align}
I_{\pm}^{>}\left(  x\right)   &  =\left(  -\sqrt{b^{2}+x}\mp b\right)
I_{0}^{>}\left(  x\right)  ,
\label{I>a}\\
I_{0}^{>}\left(  x\right)   &  =\frac{-\sqrt{m_{B}/8}}{\sqrt{b^{2}+x}%
\sqrt{b+\sqrt{b^{2}+x}}},
\end{align}
and
\end{subequations}
\begin{subequations}
\label{I<}%
\begin{align}
I_{\pm}^{<}\left(  x\right)   &  =\left[  \sqrt{-x}+\left(  1\mp1\right)
b\right]  I_{0}^{<}\left(  x\right)  ,
\label{I<a}\\
I_{0}^{<}\left(  x\right)   &  =\frac{\sqrt{b+\sqrt{b^{2}+x}}-\sqrt
{b-\sqrt{b^{2}+x}}}{\sqrt{8/m_{B}}\sqrt{-x}\sqrt{b^{2}+x}},
\end{align}
with
\end{subequations}
\begin{equation}
b=n_{B}g_{BB}.
\end{equation}

The second column of Fig.\ \ref{Fig:F(w^2)} shows how isolated roots change as a function of
$\bar{g}_{IB}$.  In Fig.\  \ref{Fig:F(w^2)}(b), which illustrates phase I, there is no isolated root.  The isolated
negative root can be seen to emerge as $\bar{g}_{IB}$ decreases from being
positive in phase I to being negative in phase II, as shown in Fig.\ \ref{Fig:F(w^2)}(d).  The isolated root then
changes from negative to positive as $\bar{g}_{IB}$ decreases further and we go
from phase II to phase III, as shown in Fig.\ \ref{Fig:F(w^2)}(f). All of this is in complete agreement with our previous analysis of $F(  w^{2})  $ using discrete sums.

We now look more closely at the discrete roots in phases II and III. In
phase II, the discrete root has $w^{2}<0$, making $w$
purely imaginary, as explained previously.
The value of $w$ is found, with the help of Eq.\ (\ref{I<a}), to obey
\begin{align}
0  &  =g_{IB}^{-2}+2I_{0}^{<}\left(  w^{2}\right)  \left(  b+\sqrt{-w^{2}%
}\right) 
\nonumber\\
&\qquad \times  \left[  \sqrt{-w^{2}}I_{0}^{<}\left(  w^{2}\right)  +g_{IB}^{-1}\right]  .
\end{align}
which can be solved analytically with the result%
\begin{equation} \label{w II}
w=\pm i\frac{\left\vert g_{IB}\right\vert m_{B}}{2}\sqrt{g_{c}^{2}-g_{IB}^{2}%
}.
\end{equation}
As displayed in Fig.\ \ref{Fig:bindingEnergy}, the (absolute) imaginary part of the root reaches its
maximum $m_{B}g_{c}^{2}/4$ at $g_{IB}=g_{c}/\sqrt{2}$ but becomes zero near the
two critical points, which are boundaries with phases I and III, where the
dynamics is expected to undergo a critical slowing-down. Because $m_{B}g_{c}^{2}/4=\gamma n_{B}^{2}/m_{B}$, for a fixed boson number
density, the smaller $\gamma$, the smaller the imaginary part.
\begin{figure}
\centering
\includegraphics[
width=3.2in
]%
{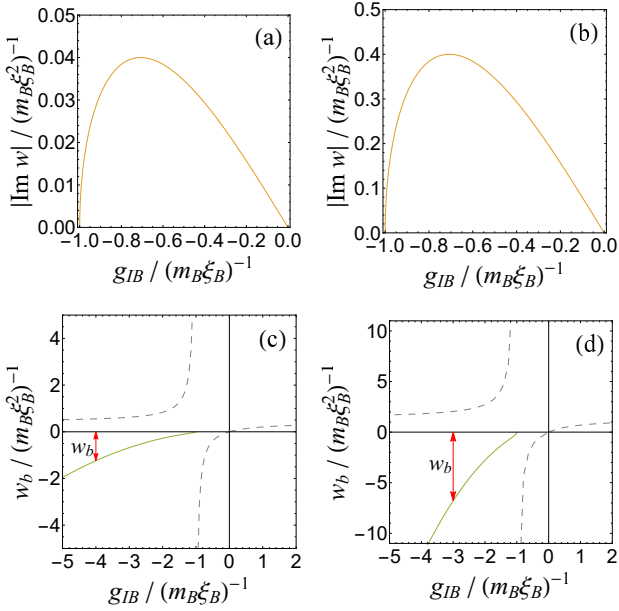}%
\caption{The top row shows the (absolute) imaginary part of the (purely imaginary) eigenvalue in Eq.\ (\ref{w II}) in phase II as a function of
$g_{IB}$ when (a) $\gamma=0.04$ and (b) 0.4. The bottom row shows
the bound-state energy, i.e.\ the single negative eigenvalue $w_{b}=-| w_b|$, where $|w_b|$ is given in Eq.\ (\ref{w_b}), in phase III as a function of $g_{IB}$ when (c) $\gamma=0.04$ and (d) 0.4 as the solid green curves.  We include in the bottom row the exact polaron energy, $E_0$, shown in Figs.\ \ref{Fig:polaronEnergy}(a) and (b), as the dashed curves.}%
\label{Fig:bindingEnergy}%
\end{figure}

A comment is in order concerning the (attractive polaron) state in phase
II always being dynamically unstable, irrespective of the coupling strength $\gamma$.  It is well known that bosons, when subject to an attractive interaction
with an impurity, have a tendency to collapse around the impurity.  For small
$\gamma$, the boson-boson repulsion may not be sufficiently strong to prevent such
collapse, but as $\gamma$ increases, the repulsion is expected to increase the stability of
the attractive polaron.  This trend appears to be confirmed by the quantum Monte
Carlo simulation by Parisi and Giorgini \cite{parisi17PhysRevA.95.023619}, who
studied a 1D Bose polaron system with a mobile impurity and a Bose gas having
a finite number of bosons.  That our results do not support this may be an
artifact of Bogoliubov theory where the repulsive phonon-phonon
interaction is ignored so that an increase in $\gamma$ will not translate
into an increased repulsion between phonons in the cloud surrounding the
impurity.  Also, for a mobile impurity, when moving to a frame attached to
the impurity, one can easily see that the impurity motion induces a
phonon-phonon interaction, which seems also to stabilize the attractive
polaron judging from the results of Grusdt et al.\
\cite{grusdt17NewJournalOfPhysics.19.103035}.

In phase III, the discrete root has $w^{2}>0$ and is found, with the help of
Eq.\ (\ref{I>a}), to obey
\begin{equation}
g_{IB}^{-1}-2\sqrt{b^{2}+w^{2}}I_{0}^{>}\left(  w^{2}\right)  =0,
\label{g_IB I_0}%
\end{equation}
which can be solved analytically with the result
\begin{equation} 
\left\vert w\right\vert =\left\vert w_{b}\right\vert \equiv\frac{\left\vert
g_{IB}\right\vert m_{B}}{2}\sqrt{g_{IB}^{2}-g_{c}^{2}}. \label{w_b}%
\end{equation}
The subscript $b$ in $w_{b}$ is to stress that this is the binding energy of a
bound state, which we now explain.%

Recall from Sec.\ IV that for every real eigenvalue, there corresponds a
partner eigenvalue with opposite sign, but only the eigenvalue whose
eigenvector has positive norm (with respect to metric $\zeta$) is physically meaningful
and is retained in calculations. To determine whether it is $w=+\left\vert
w_{b}\right\vert $ or $-\left\vert w_{b}\right\vert $ that is physically
meaningful, we first use Eqs.\ (\ref{X+X-}) to evaluate $U_{\mathbf{k}%
}-V_{\mathbf{k}}$ and $U_{\mathbf{k}}+V_{\mathbf{k}}$, which simplify, under
 condition (\ref{g_IB I_0}), to
\begin{equation}
U_{\mathbf{k}}-V_{\mathbf{k}}=\frac{2\chi_{\mathbf{k}}w_{b}}{w_{b}^{2}%
-\omega_{\mathbf{k}}^{2}}\frac{f}{\sqrt{\mathcal{V}}}\left(  1-\frac
{\omega_{\mathbf{k}}^{2}}{\epsilon_{\mathbf{k}}}\frac{2}{g_{IB}^{2}m_{B}%
}\right) \label{U_k - V_k}%
\end{equation}
and%
\begin{equation}
U_{\mathbf{k}}+V_{\mathbf{k}}=\frac{2\chi_{\mathbf{k}}\omega_{\mathbf{k}}%
}{w_{b}^{2}-\omega_{\mathbf{k}}^{2}}\frac{f}{\sqrt{\mathcal{V}}}\left(
1-\frac{2w_{b}^{2}}{\epsilon_{\mathbf{k}}g_{IB}^{2}m_{B}}\right).
\label{U_k + V_k}%
\end{equation}
Further algebraic manipulation of the product of Eqs.\ (\ref{U_k - V_k}) and
(\ref{U_k + V_k}) and the substitution of $w_{b}^{2}$ in Eq.\ (\ref{w_b}) into
the parentheses in Eq.\ (\ref{U_k + V_k}) yields 
\begin{align}
U_{\mathbf{k}}^{2}-V_{\mathbf{k}}^{2}  &  =-w_{b}\frac{f^{2}}{\mathcal{V}%
}\frac{8}{m_{B}g_{IB}^{2}}  \frac{\left[  \epsilon_{\mathbf{k}}-\frac{m_{B}}{2}\left(  g_{IB}^{2}%
-g_{c}^{2}\right)  \right]  ^{2}}{\left(  w_{b}^{2}-\omega_{\mathbf{k}}%
^{2}\right)  ^{2}},
\end{align}
which always has the opposite sign as $w_{b}$ (regardless of the momentum
mode). We thus conclude that the physical solution corresponds to the bound
state with negative energy $-| w_{b}| $. 

In one dimension, it is well known that for an atom in a negative delta function
potential $-|g_{IB}| \delta(  x)$, there
exists a single bound state with energy $-g_{IB}^2 m_{B}/2$, which, for a
mobile impurity, becomes the energy of a dimer, $-g_{IB}^{2}m_{r}/2$, where
$m_{r}$ is the reduced mass between the impurity and the atom.  It is thus
not surprising that $w_{b}$ in Eq.\ (\ref{w_b}) approaches this energy in the
limit where bosons in the bath do not interact with each other ($g_{BB}
=0=g_{c}$).  Figure \ref{Fig:bindingEnergy} illustrates the bound state energy, Eq.\ (\ref{w_b}),
as a function of $g_{IB}$.  As can be seen, the state is deeply bound
when $g_{IB}$ is tuned far less than $g_{c}$ and becomes shallowly bound when
$g_{IB}$ is tuned close to $g_{c}$ from below.  The mean-field polaron
energy develops a singularity when the bound-state energy is tuned right at
and therefore is on resonance with the continuum threshold, reminiscent of
Feshbach resonance, which occurs when a bound state in the closed channel
drops below the continuum of the ground state.

In summary, the energy spectrum in phase I consists only of the continuum and
phase I is thermodynamically stable and supports stable ground polaron states.
In phase II, there is a pair of imaginary eigenvalues and states in phase II
are dynamically unstable. Finally, the energy spectrum in phase III contains
a bound state isolated from the continuum and phase III is thermodynamically
unstable and therefore supports metastable polaron states.

\begin{figure*}
\centering
\includegraphics[
width=7in
]%
{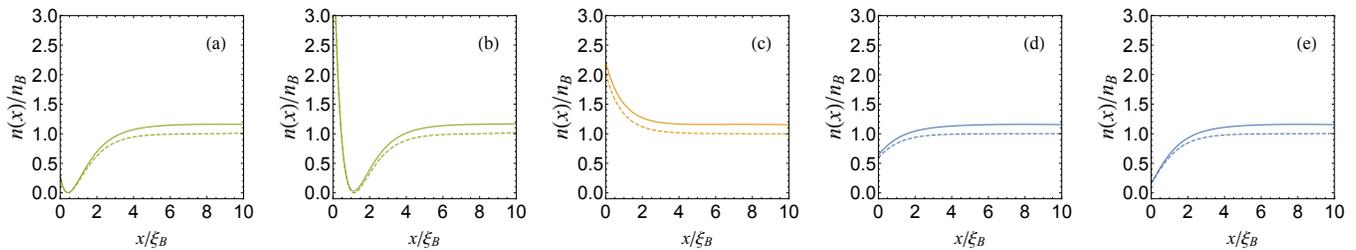}%
\caption{Boson number density profiles for (a) $g_{IB}/(m_B\xi_B)^{-1} = -3$, (b) $-1.5$, (c) $-0.3$, (d) $0.3$, and (e) $1.5$.  For all plots $\gamma = 0.4$.  (a) and (b) are in phase III, (c) is in phase II, and (d) and (e) are in phase I.}
\label{Fig:boson number density profile}%
\end{figure*}
We conclude this section by taking a look at the boson number density profile
in each phase, with the detailed derivation being left to Appendix A.  Figure
\ref{Fig:boson number density profile} samples the density profile, $n\left(  x\right)  $,
for a quasi-1D Bose gas along the $x$ dimension. In phase I where $g_{IB}>0$,
the ground state is a repulsive polaron $\left(  E_{0}>0\right)  $ and indeed
the impurity (located at the origin $x=0$) repels nearby bosons, creating a
hole at its location.  As $g_{IB}$ drops from positive to negative, the
system enters phase II, where the polaron becomes attractive ($E_{0}<0$) and
indeed the impurity pulls nearby bosons towards it, causing bosons to pile up at the impurity location.  As $g_{IB}$ is lowered below $g_{c}$, the system enters phase III, where the state becomes a
repulsive polaron again $\left(  E_{0}>0\right)$, but with a more intriguing
density profile.  In phase III, in addition to a peak at the impurity
location, a hole near (but not exactly at) the impurity location develops.
 As $g_{IB}$ gets farther away from the critical point $g_{c}$, the peak at
$x=0$ continuously decreases while the hole becomes deeper and gets close to
the impurity. As such, at the extreme end of phase III, where $g_{IB}=-\infty$,
the density profile looks identical to that at the extreme end of phase I
where $g_{IB}=+\infty$, demonstrating once again that there is a smooth
crossover between the two repulsive phases, as we saw in the polaron energy
diagram in Fig.\ 1.


\section{Conclusion}

We considered a 1D Bose polaron in the static limit where the Bose gas is
described within Bogoliubov theory but the impurity-phonon coupling is
modeled by terms that go beyond the Fr\"{o}hlich paradigm. We diagonalized exactly the beyond-Fr\"{o}hlich Hamiltonian by applying the generalized
Bogoliubov transformation.  With our exact solution, we computed boson number
density profiles and a polaron energy that is free of divergences common
to 1D systems. In Appendix \ref{app:quasiparticle} we use the quasiparticle residue as an example
to illustrate that other polaron properties can also be obtained exactly
within our approach.  Following a detailed stability analysis, we constructed
analytically the polaron phase diagram.
We found that the repulsive polaron on the negative side of the
impurity boson interaction is always thermodynamically unstable due to
the presence of a bound state existing slightly above the vacuum dimer
energy, whereas the attractive polaron
on the negative side of the impurity-boson interaction has a pair of imaginary
energies and is therefore always dynamically unstable.

A vast number of problems in many-body physics are too complex to undergo
exact treatment.  Exactly solvable models, though small in number, are
extremely useful. A good grasp of an exactly solvable model helps one to gain
insight into, and therefore find ways to solve approximately, more complicated
models that can be reduced to the exactly solvable one under limited (but
often extreme) conditions.  There are many ways in which the present study can
benefit research in Bose polarons in cold atoms.  Our exact treatment can be
adapted straightforwardly to static models in higher dimensions.  The ideas
behind our exact ansatz can be explored for developing variational ansatzes
that approximate, more accurately, the ground states of mobile polarons.  The
present study provides insight for developing and understanding more realistic
models where phonon-phonon interactions are included.  When equipped with the
local density approximation, our approach can be adapted to treat
polarons in traps of sizes much larger than the healing length.


\appendix

\section{Boson number density profiles}

The boson number density in position space, $n(  \mathbf{r})  $, is
defined as%
\begin{equation}
n\left(  \mathbf{r}\right)  =\left\langle \hat{\Psi}^{\dag}\left(
\mathbf{r}\right)  \hat{\Psi}\left(  \mathbf{r}\right)  \right\rangle,
\end{equation}
where $\hat{\Psi}(  \mathbf{r})  =\sum_{\mathbf{k}}\hat
{b}_{\mathbf{k}}^{\prime}e^{i\mathbf{k}\cdot\mathbf{r}}/\sqrt{\mathcal{V}}$ is
the boson field operator in position space. After singling out the condensed
mode, we express the number density in terms of the noncondensed field modes
$\hat{b}_{\mathbf{k}\neq0}^{\prime}$ as
\begin{align}
n\left(  \mathbf{r}\right)   &  =n_{B}+\sqrt{\frac{n_{B}}{\mathcal{V}}}%
\sum_{\mathbf{k}}\left(  \left\langle \hat{b}_{\mathbf{k}}^{\prime
}\right\rangle e^{i\mathbf{k}\cdot\mathbf{r}}+h.c\right)  \nonumber\\
& \qquad +\frac{1}{\mathcal{V}}\sum_{\mathbf{k},\mathbf{k}^{\prime}}\left\langle
\hat{b}_{\mathbf{k}}^{\prime\dag}\hat{b}_{\mathbf{k}^{\prime}}^{\prime
}\right\rangle e^{-i\left(  \mathbf{k}-\mathbf{k}^{\prime}\right)
\cdot\mathbf{r}}.
\label{n(r)}%
\end{align}
With the help of Eqs.\ (\ref{phonon field operator}) and
(\ref{shifted field operator}), we write $\hat{b}_{\mathbf{k}}^{\prime}$ in
terms of the shifted phonon field operator $\hat{c}_{\mathbf{k}}$ as%
\begin{equation}
\hat{b}_{\mathbf{k}}^{\prime}=\left(  u_{\mathbf{k}}z_{\mathbf{k}%
}-v_{\mathbf{k}}z_{\mathbf{k}}\right)  +\left(  u_{\mathbf{k}}\hat
{c}_{\mathbf{k}}-v_{\mathbf{k}}\hat{c}_{-\mathbf{k}}^{\dag}\right)  ,
\end{equation}
where $u_{\mathbf{k}}$ and $v_{\mathbf{k}}$ are defined in Eqs.\ (\ref{uv}).
It then becomes an easy exercise to show that the expectation value of
$\hat{b}_{\mathbf{k}}^{\prime}$ $\ $and $\hat{b}_{\mathbf{k}}^{\prime\dag}%
\hat{b}_{\mathbf{k}^{\prime}}^{\prime}$ in Eq.\ (\ref{n(r)}) are given,
respectively, by
\begin{equation}
\left\langle \hat{b}_{\mathbf{k}}^{\prime}\right\rangle =u_{\mathbf{k}%
}z_{\mathbf{k}}-v_{\mathbf{k}}z_{\mathbf{k}}\label{<b>}%
\end{equation}
and%
\begin{equation}
\left\langle \hat{b}_{\mathbf{k}}^{\prime\dag}\hat{b}_{\mathbf{k}^{\prime}%
}^{\prime}\right\rangle =\left(  u_{\mathbf{k}}z_{\mathbf{k}}-v_{\mathbf{k}%
}z_{\mathbf{k}}\right)  \left(  u_{\mathbf{k}^{\prime}}z_{\mathbf{k}^{\prime}%
}-v_{\mathbf{k}^{\prime}}z_{\mathbf{k}^{\prime}}\right)  +\left\langle
\cdots\right\rangle ,\label{<bb>}%
\end{equation}
where%
\begin{align}
\left\langle \cdots\right\rangle  &  =v_{\mathbf{k}}v_{\mathbf{k}^{\prime}%
}\delta_{\mathbf{k},\mathbf{k}^{\prime}}+u_{\mathbf{k}}u_{\mathbf{k}^{\prime}%
}\rho_{\mathbf{k}^{\prime}\mathbf{k}}-u_{\mathbf{k}}v_{\mathbf{k}^{\prime}%
}\kappa_{\mathbf{k},-\mathbf{k}^{\prime}}\nonumber\\
& \qquad -v_{\mathbf{k}}u_{\mathbf{k}^{\prime}}\kappa_{\mathbf{k}^{\prime
},-\mathbf{k}}+v_{\mathbf{k}}v_{\mathbf{k}^{\prime}}\rho_{-\mathbf{k}%
,-\mathbf{k}^{\prime}},
\end{align}
with
\begin{equation}
\rho_{\mathbf{kk}^{\prime}}=\left\langle \hat{c}_{\mathbf{k}^{\prime}}^{\dag
}\hat{c}_{\mathbf{k}}\right\rangle =\sum_{n}V_{n\mathbf{k}}V_{n\mathbf{k}%
^{\prime}}%
\end{equation}
the single-particle density matrix element and%
\begin{equation}
\kappa_{\mathbf{kk}^{\prime}}=\left\langle \hat{c}_{\mathbf{k}^{\prime}}%
\hat{c}_{\mathbf{k}}\right\rangle =\sum_{n}V_{n\mathbf{k}}U_{n\mathbf{k}%
^{\prime}}%
\end{equation}
the single-particle pair matrix element. Finally, replacing $\langle
\hat{b}_{\mathbf{k}}^{\prime}\rangle $ and $\langle \hat
{b}_{\mathbf{k}}^{\prime\dag}\hat{b}_{\mathbf{k}^{\prime}}^{\prime
}\rangle $ in Eq.\ (\ref{n(r)}) with Eqs.\ (\ref{<b>}) and (\ref{<bb>}),
respectively, we arrive at the boson density profile
\begin{align}
\frac{n\left(  \mathbf{r}\right)  }{n_{B}}  &  =\frac{n^{M}\left(
\mathbf{r}\right)  }{n_{B}}+\frac{1}{n_{B}\mathcal{V}}\sum_{\mathbf{k}%
}v_{\mathbf{k}}^{2}\nonumber\\
& +  \frac{1}{n_{B}}\sum_{n}\left[  I_{uV}^{n}\left(  \mathbf{r}\right)
^{2}+I_{vV}^{n}\left(  \mathbf{r}\right)  ^{2}-2I_{uV}^{n}\left(
\mathbf{r}\right)  I_{vU}^{n}\left(  \mathbf{r}\right)  \right]  ,
\end{align}
where%
\begin{equation}
\frac{n^M (\mathbf{r})}{n_{B}}=1+\frac{2}{\sqrt{n_{B}}%
}I\left(  \mathbf{r}\right)  +\frac{1}{n_{B}}I\left(  \mathbf{r}\right)  ^{2}%
\end{equation}
is the boson number density profile in the mean-field theory. In the above
equations, we introduced four summations:
\begin{align}
I\left(  \mathbf{r}\right)   &  =\frac{1}{\sqrt{\mathcal{V}}}\sum_{\mathbf{k}%
}z_{\mathbf{k}}\left(  u_{\mathbf{k}}-v_{\mathbf{k}}\right)  \cos\left(
\mathbf{k\cdot r}\right)  ,\\
I_{uV}^{n\text{ }}\left(  \mathbf{r}\right)   &  =\frac{1}{\sqrt{\mathcal{V}}%
}\sum_{\mathbf{k}}u_{\mathbf{k}}V_{n\mathbf{k}}\cos\left(  \mathbf{k\cdot
r}\right)  \text{ ,}\\
I_{vV}^{n}\left(  \mathbf{r}\right)   &  =\frac{1}{\sqrt{\mathcal{V}}}%
\sum_{\mathbf{k}}v_{\mathbf{k}}V_{n\mathbf{k}}\cos\left(  \mathbf{k\cdot
r}\right)  ,\\
I_{vU}^{n}\left(  \mathbf{r}\right)   &  =\frac{1}{\sqrt{\mathcal{V}}}%
\sum_{\mathbf{k}}v_{\mathbf{k}}U_{n\mathbf{k}}\cos\left(  \mathbf{k\cdot
r}\right)  ,
\end{align}
which become integrals over momentum in the thermodynamic limit. The density
profiles in Fig.\ \ref{Fig:boson number density profile} are produced using these equations for a 1D Bose gas along the $x$ dimension.


\section{Qausiparticle residue}
\label{app:quasiparticle}

Our exact solution allows observables that are of experimental interest to
be evaluated in systems described by the beyond-Fr\"{o}hlich Hamiltonian. In this Appendix we demonstrate the utility of our exact solution by
computing a quantity called the quasiparticle residue, whose definition we
review. At zero temperature, phonons are in the phonon vacuum state
$\left\vert 0\right\rangle $ in the absence of the impurity, but are in the
interacting ground state $\left\vert \phi\right\rangle $ in the presence of
the impurity. The modular square of the overlap integral, $\left\langle
0|\phi\right\rangle $, between the ground phonon state with and without
impurity scattering,
\begin{equation}
Z=\left\vert \left\langle 0|\phi\right\rangle \right\vert ^{2}, \label{Z}%
\end{equation}
is defined as the quasiparticle residue. The quasiparticle residue quantifies
the amount of bare impurity (vacuum) that remains in the interacting ground
state. 

The present problem has a fermionic analog---a localized impurity immersed in
a bath of fermionic atoms at zero temperature. In solid-state physics, a
well-known example is x-ray absorption, which creates a core hole in the midst
of conduction electrons \cite{Mahan00Book}. Treating the core hole as a
localized impurity represented by a static potential, Anderson
\cite{anderson67PhysRevLett.18.1049} studied the influence of this static
potential on a Fermi sea by computing the quasiparticle residue where state
$\left\vert 0\right\rangle $ in Eq.\ (\ref{Z}) is treated as the Fermi sea.
The result is summarized in the well-known formula%
\begin{equation}
Z=C_{0}N_{F}^{-\left(  \delta_{F}/\pi\right)  ^{2}},\label{Z fermions}%
\end{equation}
where $N_{F}$ is the total number of fermions and $\delta_{F}=-\tan
^{-1}\left(  g_{IF}\pi n_{F}\right)  $ (with $n_{F}$ the number density) is
the phase shift stemming from the $s$-wave scattering of electrons by the
impurity with strength $g_{IF}$ and $C_{0}$ is a prefactor which, though
complicated, has an analytical expression \cite{ohtaka90RevModPhys.62.929}.

The question we address in this Appendix is what is the analog of Eq.\
(\ref{Z fermions}) for our system described by the beyond-Fr\"{o}hlich
Hamiltonian? We begin by observing that the state $\left\vert \phi
\right\rangle $ can be obtained from the phonon vacuum $\left\vert
0\right\rangle $ after two successive unitary transformations.  The first
unitary transformation, $D(z)$, is defined by
\begin{align}
\hat{c}_{\mathbf{k}} &  =D(z)  \hat{b}_{\mathbf{k}}D^{\dag
}(z)  \notag\\
\hat{c}_{\mathbf{k}}^{\dag} &  =D(z)  \hat{b}_{\mathbf{k}}^{\dag
}D^{\dag}(z)  ,
\end{align}
where $D(  z)  $ is found, with the help of Eqs.\
(\ref{shifted field operator}), to be the well-known displacement operator
\begin{equation}
D(z)  =\exp\left[  \sum_{\mathbf{k}}
\left(\hat
z_{\mathbf{k}}
{b}_{\mathbf{k}}^{\dag}-z_{\mathbf{k}}^{\ast}\hat{b}_{\mathbf{k}}
\right)\right]
.\label{D(z)}%
\end{equation}
$D(z)  $ transforms $\hat{H}$, the Hamiltonian in Eq.
(\ref{H 1}), to
\begin{equation}
\hat{H}^{\prime}=D^{\dag}(z)  \hat{H}D(z)
,\label{H'}%
\end{equation}
where $\hat{H}^{\prime}$ is identical to $\hat{H}$ in Eq.\ (\ref{H quadratic})
except $\hat{c}_{\mathbf{k}}$ ($\hat{c}_{\mathbf{k}}^{\dag}$) in Eq.\
(\ref{H quadratic}) is replaced with $\hat{b}_{\mathbf{k}}$ ($\hat
{b}_{\mathbf{k}}^{\dag}$) which are now interpreted as the field operators in
the Hilbert space defined by $D(z)$.

The second unitary transformation, $S$, is defined by 
\begin{align}
\hat{d}_{n}  &  =S\hat{b}_{\mathbf{k}_{n}}S^{\dag}\\
\hat{d}_{n}^{\dag}  &  =S\hat{b}_{\mathbf{k}_{n}}^{\dag}S^{\dag},
\end{align}
which can be solved, in principle, from the Bogoliubov transformation
(\ref{Bogoliubov transformation}) [but with $\hat{c}_{\mathbf{k}}$ ($\hat
{c}_{\mathbf{k}}^{\dag}$) replaced with $\hat{b}_{\mathbf{k}}$ ($\hat
{b}_{\mathbf{k}}^{\dag}$)] in terms of operators defined in Fock space. $S$
transforms $\hat{H}^{\prime}$ in Eq.\ (\ref{H'}) to
\begin{equation}
\hat{H}^{\prime\prime}=S^{\dag}\hat{H}^{\prime}S,
\end{equation}
where $\hat{H}^{\prime\prime}$ is identical to $\hat{H}$ in Eq.\
(\ref{H d_n}) except $\hat{d}_{n}$ ($\hat{d}_{n}^{\dag}$) in Eq.\ (\ref{H d_n})
are replaced with $\hat{b}_{\mathbf{k}_{n}}$ ($\hat{b}_{\mathbf{k}_{n}}^{\dag
}$) which are now interpreted as field operators in the Hilbert space defined
by $S$.

As a result, the interacting polaron state is given by
\begin{equation}
| \phi\rangle =D(z)  S | 0\rangle ,
\end{equation}
where $S| 0\rangle $ is a normalized state given by
\cite{blaizot96QuantumTheoryBook} \newline%
\begin{equation}
S\left\vert 0\right\rangle =\frac{\exp\left(  \frac{1}{2}\sum_{\mathbf{k}%
,\mathbf{k}^{\prime}}\hat{b}_{\mathbf{k}}^{\dag}W_{\mathbf{kk}^{\prime}}%
\hat{b}_{\mathbf{k}^{\prime}}^{\dag}\right)  }{\sqrt[4]{\det\left(  U^{\dag
}U\right)  }}\left\vert 0\right\rangle ,
\end{equation}
where $W$ is a matrix defined as
\begin{equation}
W=U^{\ast-1}V^{\ast}.
\end{equation}

The quasiparticle residue $Z$ in Eq.\ (\ref{Z}) becomes the overlap between
coherent state $\left\vert -z\right\rangle $ and state $S\left\vert
0\right\rangle $: $Z=\left\vert \left\langle -z\right\vert S\left\vert
0\right\rangle \right\vert ^{2}$ or explicitly
\begin{equation}
Z=\frac{\exp\left[  \sum_{\mathbf{k}}\left(  \sum_{\mathbf{k}^{\prime}%
}z_{\mathbf{k}}W_{\mathbf{kk}^{\prime}}z_{\mathbf{k}^{\prime}}-z_{\mathbf{k}%
}^{2}\right)  \right]  }{\left\vert \det U\right\vert }, \label{Z bosons}%
\end{equation}
where the use of $\langle -z|0\rangle =\exp(  -\sum
z_{\mathbf{k}}^{2}/2)$ is made and all variables involved are assumed
to be real. Equation (\ref{Z bosons}) is the analog of Eq.\ (\ref{Z fermions})
we sought for our model.  As expected, Eq.\ (\ref{Z bosons}) simplifies to
the mean-field result $Z=\exp\left(  -\sum_{\mathbf{k}}z_{\mathbf{k}}%
^{2}\right)  $
\cite{shashi14PhysRevA.89.053617,grusdt17NewJournalOfPhysics.19.103035}, in
the absence of quantum fluctuations where $V$ is a null matrix and $U$ is an
identity matrix


%

\end{document}